\pdfoutput=1

\documentclass[11pt,twoside,a4paper,cmspaper,final,collab]{cms-tdr}
\def\svnVersion{158400}\def\cmsCernNo{2012-222}\def\cmsCernDate{\today}\def\cmsMessage{Submitted to the European Physical Journal C}

\begin{document}\cmsNoteHeader{TOP-11-016}

\hyphenation{had-ron-i-za-tion}
\hyphenation{cal-or-i-me-ter}
\hyphenation{de-vices}

\RCS$Revision: 151561 $
\RCS$HeadURL: svn+ssh://svn.cern.ch/reps/tdr2/papers/TOP-11-016/trunk/TOP-11-016.tex $
\RCS$Id: TOP-11-016.tex 151561 2012-10-09 13:02:59Z psilva $
\newlength\cmsFigWidth
\ifthenelse{\boolean{cms@external}}{\setlength\cmsFigWidth{0.95\columnwidth}}{\setlength\cmsFigWidth{0.6\textwidth}}
\ifthenelse{\boolean{cms@external}}{\providecommand{\cmsLeft}{top}}{\providecommand{\cmsLeft}{left}}
\ifthenelse{\boolean{cms@external}}{\providecommand{\cmsRight}{bottom}}{\providecommand{\cmsRight}{right}}
\ifthenelse{\boolean{cms@external}}{\providecommand{\cmsEPJCbreak[1]}{\linebreak[#1]}}{\providecommand{\cmsEPJCbreak[1]}{\relax}}

\cmsNoteHeader{TOP-11-016} 
\title{
Measurement of the top-quark mass in \ttbar\ events with dilepton final states in pp collisions at $\sqrt{s}=7\TeV$}

\date{\today}

\abstract{
The top-quark mass is measured in proton-proton collisions at $\sqrt{s} = 7\TeV$
using a data sample corresponding to an integrated luminosity of 5.0\fbinv collected by the CMS experiment at the LHC.
The measurement is performed in the dilepton decay channel
$\ttbar\rightarrow(\ell^{+}\nu_\ell \cPqb)\,(\ell^{-}\cPagn_\ell\cPaqb)$,
where $\ell=\Pe,\Pgm$.
Candidate top-quark decays are selected by requiring two leptons, at least two jets, and imbalance in transverse momentum.
The mass is reconstructed with an analytical matrix weighting technique using distributions derived from simulated samples.
Using a maximum-likelihood fit, the top-quark mass is determined to be $172.5\pm 0.4\stat \pm1.5\syst\GeV$.
}

\hypersetup{%
pdfauthor={CMS Collaboration},%
pdftitle={Measurement of the top-quark mass in t t-bar events with dilepton final states in pp collisions at sqrt(s)=7 TeV},%
pdfsubject={CMS},%
pdfkeywords={CMS, top, quark, mass}}

\maketitle 

\newcommand{\mymet}{\makebox[2.4ex]{\ensuremath{\not\!\! E_{\mathrm{T}}}}}
\providecommand{\ee}{\ensuremath{\Pe \Pe}\xspace}
\newcommand{\emu}{\ensuremath{\Pe\mu}\xspace}
\newcommand{\mumu}{\ensuremath{\mu\mu}\xspace}
\newcommand{\mtop}{\ensuremath{m_{\cPqt}}\xspace}
\newcommand{\mb}{\ensuremath{m_{\cPqb}}\xspace}

\section{Introduction}
\label{seq:introduction}

The top-quark mass is an important parameter of the standard model (SM) of particle physics, as it affects predictions of SM observables  via radiative corrections.
Precise measurements of the top-quark mass are critical inputs to global electroweak fits~\cite{LEPewkfits,globalEWKfits},
which provide constraints on the properties of the Higgs boson.

The top quark constitutes an exception in the quark sector as it decays, primarily to a W boson and a b quark, before it can hadronize.
Thus, in contrast to all other quarks, the mass of the top quark can be measured directly
and is currently known with the smallest relative uncertainty.
All measurements of the top-quark mass to date are based on the decay products of \ttbar\ pairs, using final states with zero, one, or two charged leptons.
The mass of the top quark has been measured very precisely in  $\Pp\bar\Pp$ collisions by the Tevatron experiments, and
the current world average is $\mtop = 173.18\pm 0.56\stat \pm 0.75\syst\GeV$~\cite{Aaltonen:2012ra}.
In the dilepton channel, in which each \PW\ boson decays into a charged lepton and a neutrino,
the top-quark mass has been measured to be
$\mtop = 170.28\pm 1.95\stat \pm 3.13\syst\GeV$ by the CDF Collaboration~\cite{Aaltonen:2011dr} and
$\mtop = 174.00\pm 2.36\stat \pm 1.44\syst\GeV$ by the D0~Collaboration~\cite{Abazov:2012rp}.
The combination of these two measurements yields a top-quark mass of
$\mtop = 171.1\pm 2.1\GeV$~\cite{Aaltonen:2012ra}.
Measurements of $\mtop$ in $\cmsSymbolFace{p}\cmsSymbolFace{p}$ collisions at $\sqrt{s}=7$~TeV
were performed at the Large Hadron Collider (LHC) in the dilepton channel
by the Compact Muon Solenoid (CMS) Collaboration~\cite{Chatrchyan:2011nb} and in the lepton+jet channel,
in which one \PW\ boson decays into quarks and the other into a charged lepton and a neutrino,
by the ATLAS~\cite{ATLAS:2012aj} and CMS~\cite{CMS-PAS-TOP-11-015} Collaborations.

Of all \ttbar decay channels, the dilepton channel has the smallest branching fraction
and is expected to be the least contaminated by background processes.
The dominant background process is Drell--Yan (DY) production. Single top quark production through the $\cPqt\PW$ channel as well as
diboson production also mimic the dilepton signature but have much lower cross sections.
The production of multijet events has a large cross section at the LHC,
but the contamination of the dilepton sample is small
as two isolated leptons with high transverse momentum ($\pt$) are very rarely produced.
The presence of at least two neutrinos in dilepton \ttbar decays gives rise to an experimental $\pt$ imbalance,
which allows a further discrimination between background and \ttbar events.
However, the kinematical system is underconstrained as only the $\pt$ imbalance can be measured.

Here we report an update of the measurement of $\mtop$
performed in dileptonic final states, containing electrons or muons,
with an analytical matrix weighting technique.
An alternative measurement is performed using a full kinematic analysis.
The data samples used in this analysis were recorded by the CMS experiment 
at a centre-of-mass energy of $7\TeV$ and correspond to a total integrated luminosity of $5.0\pm 0.1\fbinv$.

\section{The CMS detector}
\label{sec:apparatus}

The central feature of the CMS detector is a superconducting solenoid,
13\unit{m} in length and 6\unit{m} in diameter, which provides an axial magnetic
field of 3.8\unit{T}.  The bore of the solenoid is outfitted with various
particle detection systems.  Charged particle trajectories are
measured by the silicon pixel and strip subdetectors, covering $0 < \phi <
2\pi$ in azimuth and $|\eta |<2.5$, where the pseudorapidity $\eta$ is
defined as $\eta = -\ln[\tan(\theta/2)]$, with $\theta$ being the
polar angle of the trajectory of the particle with respect to the
anticlockwise-beam direction.
A lead tungstate crystal electromagnetic calorimeter (ECAL)
 and a brass/scin\-til\-lator sampling hadronic calorimeter (HCAL) surround
the tracking volume;
in this analysis the calorimetry provides high-resolution energy and
direction measurements of electrons and hadronic jets.
Muons are measured in drift tubes, cathode strip chambers, and resistive plate chambers embedded in the flux-return yoke of the solenoid.
The detector is nearly hermetic, allowing for \pt imbalance
measurements in the plane transverse to the beam directions.
A two-level trigger system selects the most interesting pp collision
events for use in physics analysis.
A detailed description of
the CMS detector can be found in Ref.\cite{cms}.

\section{Simulation of signal and background events}
\label{sec:samples}

The simulation of \ttbar
events is performed using the \MADGRAPH~\cite{MADGRAPH}
event generator (v.~5.1.1.0),
where the generated top-quark pairs are accompanied by up to three
additional high-\pt jets. The parton configurations generated by \MADGRAPH are processed with
\PYTHIA~6.424~\cite{Sjostrand:2006za} to provide showering of the generated particles.
The parton showers are matched  using the $\kt$-MLM prescription~\cite{mlm}.
The underlying event is described with the \textsc{Z2} tune~\cite{Chatrchyan:2011id}
and the CTEQ6.6L~\cite{Nadolsky:2008zw} set of parton distribution functions (PDFs) are used.
The \TAUOLA package (v.~27.121.5) \cite{Davidson:2010rw} is used to simulate decays of the $\tau$ leptons.
Events in which the $\tau$ leptons decay to electrons or muons are taken as part of the signal.

For the reference sample, a top-quark mass of $\mtop = 172.5\GeV$ is used. Additional samples with masses of \cmsEPJCbreak{4} 161.5\GeV and between
163.5 and 187.5\GeV in steps of 3\GeV are used.
Furthermore, in order to estimate systematic effects in the modelling of dilepton events,
simulated signal samples using alternative settings of the parameters are also considered.
The following parameters are varied:
the QCD factorisation and renormalisation scale (defined
as the squared sum of the four-momenta of the primary partons in the event which is transferred dynamically in the hard interaction)
and the threshold used for the matching of the partons from matrix elements to the parton showers.
The uncertainty on the choice of the $Q^2$ or matching scales are considered by varying the corresponding nominal value by a factor of two, up and down.

Electroweak production of single top quarks is simulated using \POWHEG~(v.~301)~\cite{powheg};
\MADGRAPH is used to simulate $\PW/\Z$ events with up to four jets.
Production of \PW\PW, \PW\Z, and $\Z\Z$ is simulated with \PYTHIA.

Signal and background processes used in the analysis of \ttbar events are normalised to
next-to-leading order (NLO) or next-to-next-to-leading order (NNLO) cross section calculations, where calculations are available.
The production cross section of $\sigma_{\ttbar}=164^{+13}_{-10}$\unit{pb} computed with \textsc{HATHOR}~\cite{Langenfeld:2009wd,Aliev:2010zk} at approximate NNLO is used.
The single top quark associated production (\cPqt\PW) cross section is taken to be $\sigma_{\cPqt\PW}=15.7\pm1.2\unit{pb}$ at NNLO~\cite{Kidonakis:2010tW}.
The inclusive NNLO cross section of the production of \PW\ bosons
(multiplied by the leptonic branching fraction of the \PW\ boson)
is estimated to be  $\sigma_{\PW\rightarrow \ell\nu} = 31.3 \pm 1.6\unit{nb}$ using \textsc{fewz}~\cite{fewz}
with a $Q^2$ scale of $(m_{\PW})^2 + \sum (\pt^\text{parton})^2$, where $m_\PW=80.4\GeV$
and $\pt^\text{parton}$ are the transverse momenta of the partons in the event.
The DY production cross section at NNLO is calculated using \textsc{fewz}
to be $\sigma_{\Z/\gamma^*\rightarrow \ell\ell} (m_{\ell\ell}>20\GeV) = 5.00 \pm 0.27$\unit{nb},
where $m_{\ell\ell}$ is the invariant mass of the two leptons.
In the computation the scales are set using the \Z-boson mass $m_\Z=91.2\GeV$~\cite{PDG}.
The normalisation of \PW\PW, \PW\Z, and $\Z\Z$ production is defined using the inclusive cross sections
of $43.0\pm1.5$\unit{pb}, $18.8\pm 0.7$\unit{pb}, and $7.4\pm 0.2$\unit{pb} respectively
(all calculated at NLO with \MCFM~\cite{Campbell:2011bn}).

All generated events are passed through the full simulation of the CMS detector based on \GEANTfour~\cite{GEANT4}.
We simulate additional soft Monte Carlo events corresponding to a number of collisions distributed as seen in data.

\section{Event selection}
\label{sec:evtsel}

The \ttbar candidate events are required to contain at least two jets,
two energetic isolated leptons (electrons or muons), and missing transverse energy (\MET)
which is defined as the magnitude of the \pt imbalance vector.
Events are selected by dilepton triggers in which two muons, two electrons,
or one electron and one muon are required to be present.
The instantaneous luminosity increased significantly during the data taking period thus
the lepton $\pt$ thresholds were increased during the data taking period to keep the trigger rates within the capabilities of the data acquisition system.
For the dimuon trigger, the \pt requirements evolved from 7\GeV for each muon to  asymmetric requirements of 17\GeV for the highest-\pt (leading) muon and 8\GeV for the second-highest \pt muon.
For the dielectron trigger, the requirement was asymmetric with a threshold
applied to the energy of an ECAL cluster projected onto the plane transverse to the nominal beam line ($\ET$).
The cluster of the leading electron is required to have $\ET>17\GeV$ and the second-leading electron $\ET>8\GeV$.
For the electron-muon trigger, the
thresholds were either $\ET>17\GeV$ for the electron and $\pt>8\GeV$ for the muon, or $\ET>8\GeV$ for the electron
and $\pt>17\GeV$ for the muon.

All objects are reconstructed using a particle-flow  algorithm~\cite{CMS-PAS-PFT-10-002}.
The particle-flow algorithm combines the information from all subdetectors to identify and
reconstruct all particles produced in the collision, namely charged hadrons, photons, neutral
hadrons, muons, and electrons.
Jets are reconstructed by the anti-$\kt$ jet clustering algorithm \cite{Cacciari:2008gp} with a distance parameter $R = 0.5$.
Jet energy corrections are applied to all the jets in data and simulation~\cite{Chatrchyan:2011ds}.
The \MET vector is calculated using all reconstructed particles.

Events are selected with two isolated, oppositely charged leptons
with $\pt>20\GeV$ and $|\eta|<2.4$, and
at least two jets with $\pt>30\GeV$ and $|\eta|<2.4$.
The lepton isolation $I_{\text{rel}}$ is defined as the sum of the transverse momenta of 
stable charged hadrons, neutral hadrons and photons in a cone of $\Delta R = \sqrt{(\Delta\eta)^2+(\Delta\phi)^2} = 0.3$ around the lepton track,
divided by its transverse momentum.
A lepton candidate is not considered as isolated and is rejected if the value of $I_{\text{rel}}$ is $>0.20$ for a muon and $>0.17$ for an electron.
The two leptons of highest \pt are chosen for the reconstruction of the top quark candidates. The choice of the jets is different in each analysis and is described later.
The reconstructed \MET of events with same-flavour lepton pairs is required to be above 40\GeV to reject DY events.
No such selection is applied to $\Pe \mu$ events.
The selected leptons and jets are required to originate from the primary $\Pp\Pp$ interaction vertex, identified as the reconstructed vertex with the largest $\sum\pt^2$ of its associated tracks.
Events with same-flavour lepton pairs in the dilepton mass window $76<m_{\ell\ell}<106\GeV$ are removed to suppress the dominant DY production background.
Dilepton pairs from heavy-flavour resonances as well as low-mass DY production are also removed by requiring a minimum invariant mass of 20\GeV.
A highly efficient \cPqb-tagging algorithm based on a likelihood method that combines information
about impact parameter significance, secondary vertex reconstruction, and jet kinematic properties,
into a \cPqb-tagging  discriminator,
is used to classify the jets~\cite{CMS-PAS-BTV-11-004}.
We require at least one \cPqb-tagged jet in the event.

The observed number of events is consistent with the expected signal and background yields, as shown in Table~\ref{tab:CutFlow_DL}.
Simulated events are reweighted to account for differences in trigger, lepton, and b-tagging  selection efficiency between data and simulation.
The \cPqb-tagging efficiency is estimated from a sample of top-quark candidates~\cite{CMS-PAS-BTV-11-003}
while the probability of tagging light-quark jets
(mistag rate) is estimated from multijet events~\cite{CMS-PAS-BTV-11-004}.
The lepton selection efficiency data-to-simulation scale factors
are estimated using dileptons inside the \Z-boson mass window.
The trigger efficiencies are estimated using a data sample collected
with a trigger based on \MET that is weakly correlated with the dilepton triggers
and after selecting dilepton events which fulfil the complete event selection criteria.

The contribution of the DY background  is measured using data.
For the \ee\ and \mumu\ channels, the $R_\text{out/in}$ method is used~\cite{Chatrchyan:2011nb}.
In this method, the number of DY events  counted inside the \Z-boson mass window in the data is rescaled by the
ratio of DY events predicted by the simulation outside and inside the mass window.
As contamination from non-DY backgrounds is expected to be present in the \Z-boson
 mass window, a  subtraction based on data is applied
using the \emu\ channel scaled
 according to the event yields in the \ee\ and $\mu\mu$ channels.
For the \emu\ channel, the DY  background yield is estimated
after performing a binned maximum-likelihood 
fit to the dilepton invariant mass distribution.
The fitting functions
are taken from simulation for both the signal and background contributions.
The  contamination from multijet and \PW+jets backgrounds is estimated with a matrix method~\cite{CMS-TOP-11-005}, and non-dileptonic \ttbar\ decays are reweighed in the simulation to take these backgrounds into account.
This component will be called {\em \ttbar\ background} in the following.

\begin{table}[t]
  \begin{center}
\topcaption{
Numbers of observed and expected events in each dilepton channel after all selection requirements have been applied.
Event yields correspond to an integrated luminosity of $5.0\fbinv$.
The uncertainties quoted correspond to the limited statistics in simulation.
The total uncertainty associated to the estimates from data of the \ttbar\ background and DY production are included as well.
}
\label{tab:CutFlow_DL}
\begin{tabular}{l|c|c|c}
\hline
Processes & ee & e$\mu$ & $\mu\mu$ \\
\hline
\hline
\multicolumn{4}{c}{1 \cPqb-tagged jet}\\
\hline
\ttbar\ signal         & 598 $\pm$ 18       & 2359 $\pm$ 71    & 770 $\pm$ 23\\
\ttbar\ background     & 10.6 $\pm$ 0.3     & 101.8 $\pm$ 3.1  & 15.7 $\pm$ 0.5\\
Single top             & 40.7 $\pm$ 1.2     & 172.2 $\pm$ 5.2  & 53.3 $\pm$ 1.6\\
Drell--Yan             & 107 $\pm$ 24       & 241 $\pm$ 27     & 143 $\pm$ 31\\
Dibosons               & 11.4 $\pm$ 0.3     & 39.7 $\pm$ 1.2   & 13.0 $\pm$ 0.4\\
\hline
Total prediction       & 767 $\pm$ 30	    & 2914 $\pm$ 76    & 995 $\pm$ 39\\
Data                   & 817                & 2788             & 1032\\
\hline\hline
\multicolumn{4}{c}{$\geq 2$ \cPqb-tagged jets}\\
\hline
\ttbar\ signal         & 1057 $\pm$ 32      & 4312 $\pm$ 129        & 1393 $\pm$ 42\\
\ttbar\ background     & 4.6 $\pm$ 0.3      & 37.6 $\pm$ 1.1        & 5.5 $\pm$ 0.5\\
Single top             & 36.8 $\pm$ 1.1     & 140.6 $\pm$ 4.2       & 48.2 $\pm$ 1.4\\
Drell--Yan             & 38 $\pm$ 11        & 38.9 $\pm$ 4.3        & 32 $\pm$ 12\\
Dibosons               & 2.9 $\pm$ 0.1      & 9.1 $\pm$ 0.3         & 2.5 $\pm$ 0.1\\
\hline
Total prediction       & 1139 $\pm$ 34      & 4539 $\pm$ 130        & 1481 $\pm$ 43\\
Data                   & 1151               & 4365                  & 1474\\
\hline
\end{tabular}
\end{center}
\end{table}

\section{Analytical matrix weighting technique}
\label{sec:amwt}

Since the dilepton channel contains in the final state at least two neutrinos which can not be detected,
the reconstruction of $\mtop$ from dilepton events involves an underconstrained system.
For each \ttbar\ event, the kinematic properties are fully specified by 24 parameters,
which are the four-momenta
of the six particles in the final state: two charged leptons, two neutrinos and two jets.
Out of the 24 free parameters, 14 are inferred from measurements (the three-momenta of the jets and leptons, and the two components of the \MET) and 9 are constrained.
Two constraints arise from demanding that the reconstructed \PW-boson masses be equal to the world-average measured value~\cite{PDG} and
one constraint is imposed by assuming the top quark and antiquark masses to be the same~\cite{Chatrchyan:2012ub}.
Further, the masses of the 6 final-state particles are taken as the world-average measured values~\cite{PDG}.
This leaves one free parameter that must be constrained by using some hypotheses.

Several methods have been developed for measuring the top-quark mass in the dilepton decay channel. We use an improved version of the Matrix Weighting Technique (MWT)~\cite{D01998} that was used in the first measurements in this channel~\cite{D01998,CDF1997}.
The algorithm is referred to as the analytical MWT (AMWT) method.
A key improvement with respect to the original MWT is the selection of the jets used to reconstruct the top quark candidates.
Instead of taking the two leading jets (\ie\ the jets with the highest \pt) , the fraction of correctly assigned jets can be increased significantly by using the information provided by \cPqb-tagging.
Therefore, the leading \cPqb-tagged jets are used in the reconstruction, even if they are not the leading jets.
If there is a single \cPqb-tagged jet in the event, it is supplemented by the leading untagged jet.
The same \cPqb-tagging algorithm is used as in the event selection.
A further improvement is the use of an analytical method~\cite{PhysRevD.73.054015, *PhysRevD.78.079902}  to determine the momenta of the two neutrinos instead of a numerical method.

In the AMWT, the mass of the top quark is used to fully constrain the \ttbar\ system.
For a given top-quark mass hypothesis,
the constraints and the measured observables restrict the transverse momenta of the neutrinos to lie on ellipses
in the $p_x$-$p_y$ plane.
If we assume that the measured missing transverse energy is solely due to the neutrinos,
the two ellipses constraining the  transverse momenta of the neutrinos can be obtained, and
the intersections of the ellipses provide the solutions that fulfill the constraints.
With two possible lepton-jet combinations, there are up to eight solutions for the neutrino momenta for a given top-quark mass hypothesis.
Nevertheless, in this method, an irreducible singularity that precludes the determination of the longitudinal momentum of the neutrinos remains in a limited kinematical region.
The fraction of events affected by this singularity is 
below  0.1\%,
and a numerical method is used to determine the solutions in these rare cases~\cite{PhysRevD.72.095020}.

The kinematic equations are solved many times per event using a series of top-quark mass hypotheses between 100 and $400\GeV$ in $1\GeV$ steps.
Typically, solutions are found for the neutrino momenta that are consistent with all constraints for large intervals of mass hypotheses.
In order to determine a preferred mass hypothesis, a weight $w$ is assigned to each solution~\cite{DG}:
\begin{equation}
w = \left\{\sum f(x_1)f({x_2})\right\}p(E_{\ell^+}^*|\mtop)p(E_{\ell^-}^*|\mtop)\;,
\label{eq:evweight}
\end{equation}
where $x_i$ are the Bjorken $x$ values of the initial-state partons, $f(x)$ are the parton distribution functions, and the summation is over the possible leading-order initial-state partons
($\cPqu\cPaqu$, $\cPaqu\cPqu$, $\cPqd\cPaqd$, $\cPaqd\cPqd$, and $\Pg\Pg$).
Each term of the form $p(E^*|\mtop)$ is the probability density of observing a massless charged lepton
of energy $E^*$ in the rest frame of the top quark, for a given $\mtop$~\cite{DG}:
\begin{equation}
p(E^*|\mtop)=\frac{4\mtop E^*(\mtop^2-\mb^2-2\mtop E^*)}{(\mtop^2-\mb^2)^2+M_W^2(\mtop^2-\mb^2)-2M_W^4}\;.
\label{eq:probdens}
\end{equation}

Detector resolution effects are accounted for by
reconstructing the event 1000 times, each time varying the
$\pt$, $\eta$, and $\phi$ of each jet according to the measured detector resolution, and correcting the \MET accordingly.
For each mass hypothesis, the weights $w$ from all solutions are summed.
For each event, the top-quark mass hypothesis with the maximum weight is taken as the reconstructed top-quark mass $m_{\rm AMWT}$.
Events that have no solutions or that have a maximum weight below a threshold are discarded.
This removes 14.6\% of the events, and
 9934 events remain in the data, 1550 \ee events, 6222 \emu events, and 2110 \mumu events.

\begin{figure}[htbp]
  \begin{center}
  \includegraphics[width=0.45\textwidth]{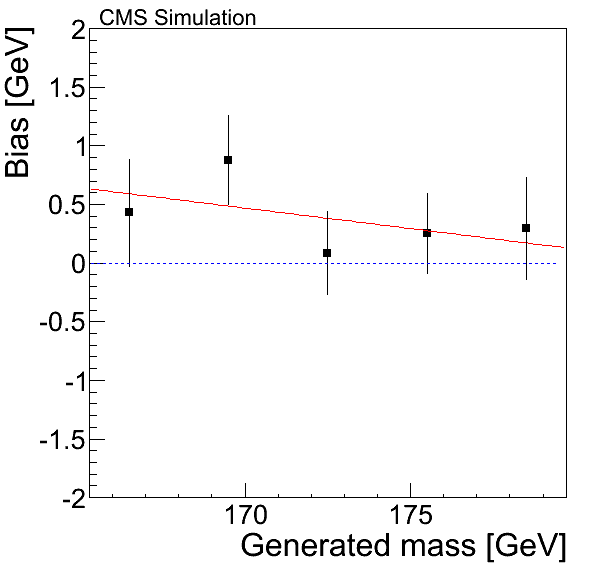}
 \includegraphics[width=0.45\textwidth]{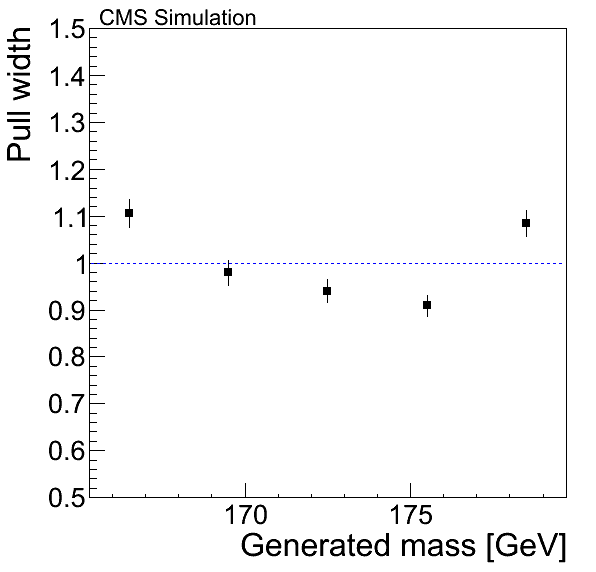}
    \caption{Mean mass bias (\cmsLeft) and pull width (\cmsRight) for different top-quark masses in pseudo-experiments for the AMWT method.
    The red solid line represents the linear fit used to determine the correction to apply in order to minimise the residual bias and the blue dashed line show the expectation for an unbiased fit.
    The average pull width for the different top-quark masses is 0.99.
    }
    \label{fig:calib_B_tag}
  \end{center}
\end{figure}

A likelihood $\cal L$ is computed for values of $\mtop$ between 161.5 and 184.5\GeV,
from data in the range $100<m_\mathrm{AMWT} \cmsEPJCbreak{3} <300\GeV$.
For each value of $\mtop$, the likelihood is computed by 
comparing the reconstructed mass distribution in data with the expectation from simulation. For the background, the reconstructed mass distribution
of each individual process is added according to its expected relative contribution.
Two different templates are used according to the \cPqb-tag multiplicity of the event, either one \cPqb-tagged jet,
or two or more \cPqb-tagged jets.
For the DY background, the relative contribution is derived from data in the \Z-boson mass window.
For the other processes, the contributions predicted by the simulation are used.
The value that maximises the likelihood is calculated after fitting a quadratic function to
the $-\ln\mathcal{L}$ values obtained for all mass points and it
is taken as the measurement of $\mtop$.
Using all the mass points in this fit yields pull widths that are closer to unity.

We determine the bias of this estimate using ensembles of pseudo-experiments based on the expected numbers of signal and background events, as shown in Fig.~\ref{fig:calib_B_tag}.
Given the fit to the data, a correction of $-0.34\pm0.20\GeV$ is applied to the final result to compensate for the residual bias introduced by the fit (Fig.~\ref{fig:calib_B_tag}, \cmsLeft).
This correction is obtained from the fit of a linear function to the average top-quark masses measured for different mass hypotheses.
The width of the pull distribution is within 10\% of unity for all the mass points,
indicating that the statistical uncertainties are correctly estimated (Fig.~\ref{fig:calib_B_tag}, \cmsRight).

\begin{figure}[t]
  \begin{center}
        \includegraphics[width=\cmsFigWidth]{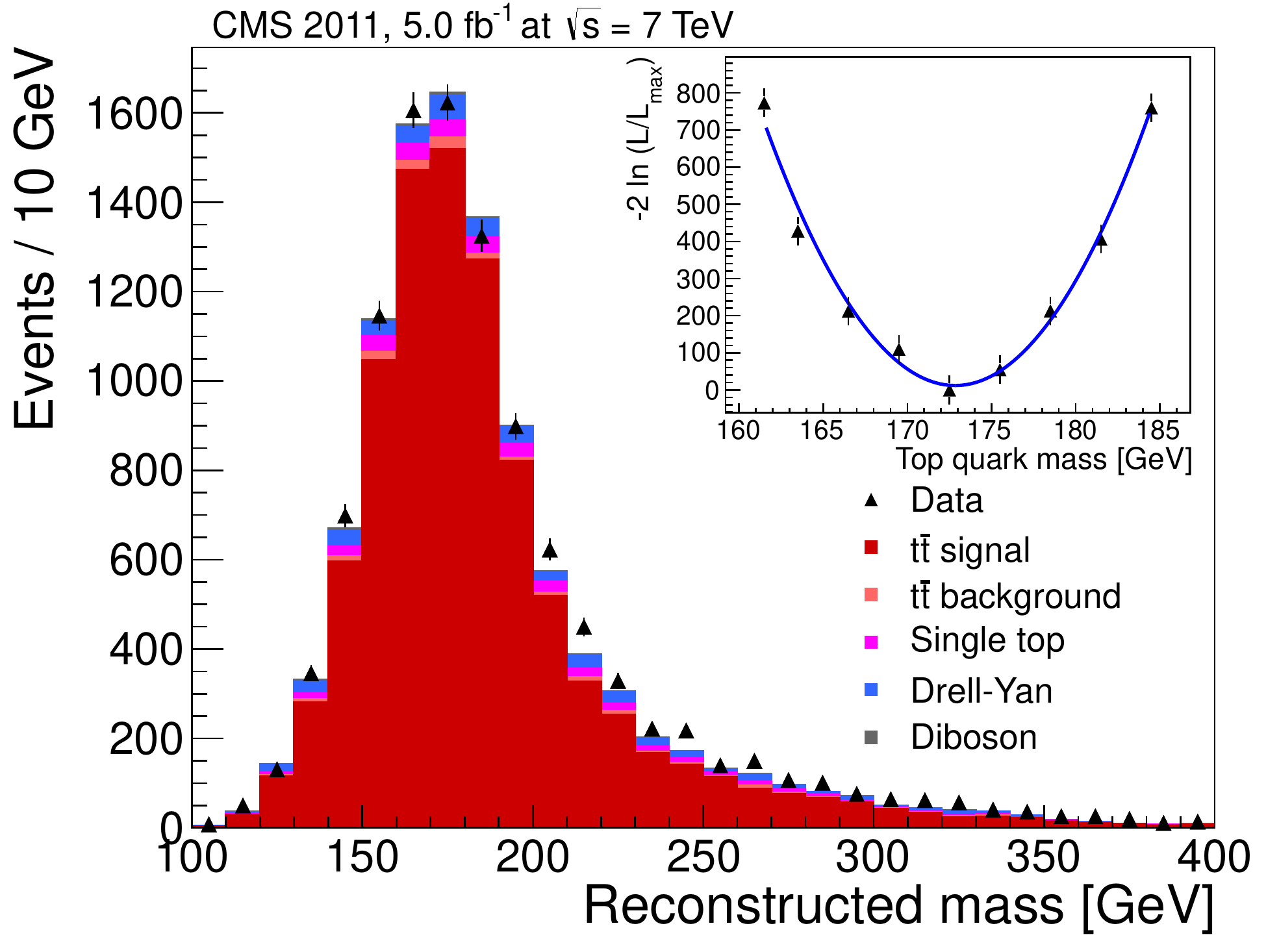}
        \caption{Distribution of the reconstructed mass in data and simulation for a top-quark mass hypothesis of 172.5\GeV
with the AMWT method.
All events used in the analysis are included in the distribution.
The inset shows $-2\ln(\mathcal{L}/\mathcal{L}_{\text{max}})$ versus $m_{\cPqt}$ with the quadratic fit superimposed.}
        \label{fit:fit_ll}
  \end{center}
\end{figure}

After correction for the bias, the top-quark mass is measured to be $\mtop=172.50\pm 0.43\stat\GeV$.
The predicted distribution of the reconstructed masses $m_\mathrm{AMWT}$ for a simulated top quark with mass $\mtop = 172.5\GeV$,
superimposed on the distribution observed in data, is shown in Fig.~\ref{fit:fit_ll}. The inset shows the distribution of the \cmsEPJCbreak{4}  $-2\ln(\mathcal{L}/\mathcal{L}_{\text{max}})$ points with the quadratic fit used to measure $\mtop$. The $\chi^2$ probability of the fit is 0.36.

\section{Systematic uncertainties}
\label{subsec:systunc}

\begin{table}[t]
\centering
\topcaption{List of systematic uncertainties with their contributions to the top-quark mass measurement.}
\label{tab:syst_amwt}
\begin{tabular}{lc}
\hline\hline
\rule{0pt}{2.2ex}   Source &  $\Delta \mtop\ (\GeVns)$ \\
\hline
\rule{0pt}{2.2ex}Jet energy scale	& ${}^{+0.90}_{-0.97}$ \\ 
\rule{0pt}{2.2ex}b-jet energy scale      & ${}^{+0.76}_{-0.66}$ \\ 
\rule{0pt}{2.2ex}Jet energy resolution	& $\pm0.14$ \\
Lepton energy scale	& $\pm0.14$  \\
Unclustered \MET  	& $\pm 0.12$  \\
\cPqb-tagging efficiency		& $\pm0.05$  \\
Mistag rate		& $\pm0.08$ \\
Fit calibration 	& $\pm0.40$  \\
Background normalization& $\pm0.05$ \\
Matching scale		& $\pm0.19$  \\
Renormalisation and factorisation scale	& $\pm0.55$\\
Pileup			& $\pm0.11$ \\
PDFs      		& $\pm0.09$ 	  \\
Underlying event	& $\pm0.26$\\
Colour reconnection 	& $\pm0.13$  \\
Monte Carlo generator	& $\pm0.04$\\
\hline
\rule{0pt}{2.2ex}Total                   & $\pm1.48$ \\
\hline\hline
\end{tabular}
\end{table}

The contributions from the different sources of uncertainty are summarised in Table~\ref{tab:syst_amwt}.
The uncertainty of the overall jet energy scale (JES) is the dominant source of uncertainty on $\mtop$.
The JES is known with an uncertainty of 1--3\%, depending on the \pt and $\eta$ of the jet~\cite{Chatrchyan:2011ds}.
Even in a high-pileup regime such as the one observed throughout the 2011 data taking period,
the JES uncertainty is mostly dominated by the uncertainties on the absolute scale, initial- and final-state radiation, and
corrections arising from the fragmentation and single-particle response in the calorimeter.
It has been evaluated for 16 independent sources of systematic uncertainty.
To estimate the effect of each source on the measurement of $\mtop$, the ($\pt$, $\eta$)-dependent uncertainty is used to shift concurrently the energy of each jet by $\pm 1 \sigma$ with respect to its nominal value, and correcting the \MET accordingly.
For each source, pseudo-experiments are generated from simulated event samples for which the JES is varied by the relevant uncertainty,
and the reconstructed top-quark mass distributions are fitted with the templates derived with the nominal JES.
The average variation of the top-quark mass is used to estimate the systematic uncertainty.
The quadratic sum of the variation for each source is taken as the systematic uncertainty.
The uncertainty on pileup corrections to the jet energy calibration (5 sources) correspond to a combined uncertainty of 0.53\GeV on \mtop. Another important contribution is the overall data-to-simulation scale calibrated in photon+jet events, yielding a 0.51\GeV uncertainty. Other contributions are related to limited knowledge of the single-pion response ($\mbox{}^{+0.2}_{-0.3}\GeV$) and fragmentation models (0.3\GeV) used in the extrapolation as a function of jet \pt. We also include a time-dependent effect (0.2\GeV) related to variations in calorimeter response in the endcaps. Residual eta-dependent corrections based on dijet balance studies (6 sources) yield a negligible uncertainty on \mtop (0.03\GeV). All these sources added in quadrature give a combined JES uncertainty of ${}^{+0.90}_{-0.97}\GeV$. The final component of JES uncertainty corresponds to the uncertainty on the modeling of jet flavour dependence of the jet energy scale (${}^{+0.76}_{-0.66} \GeV$) which is quoted separately in Table~\ref{tab:syst_amwt}.

The uncertainty due to jet energy resolution is evaluated from pseudo-experiments where the jet energy resolution width in the simulation is modified by $\pm 1 \sigma$ with respect to its nominal width.
The uncertainty on the lepton energy scale is observed to have an almost negligible effect on the measurement of $\mtop$.
The uncertainty in the \MET scale is propagated to the measurement of $\mtop$ after subtracting
the clustered (\ie\ jet energy) and leptonic components, which are varied separately as previously described.
This procedure takes into account possible correlations between the different sources of uncertainty.
The scale of the residual unclustered energy contribution to the \MET is varied by $\pm$10\% and
the corresponding variation of the top-quark mass measurement is evaluated from pseudo-experiments.

The uncertainty due to \cPqb-tagging efficiency was evaluated by varying the
\cPqb-tagging efficiency and mistag rates of the algorithm by their respective uncertainties~\cite{CMS-PAS-BTV-11-003, CMS-PAS-BTV-11-004}.
The tagging rate was varied according to the flavour of the selected jet as determined from the simulation.
This affects the multiplicity of \cPqb-tagged jets and the choice of the jets used in the reconstruction of $\mtop$.

The effect of statistical fluctuations in the templates
is estimated by splitting the \ttbar\ sample in four independent subsamples and producing independent templates for each.
Pseudo-experiments are performed using each new signal template, and the RMS variation of the average top-quark mass from each template is taken as an estimate of this uncertainty.
The uncertainty on the calibration of the fit is added to the systematic uncertainty.
The contribution from the uncertainty in the ratio between the signal and the background used in the fit
is evaluated by varying by the corresponding uncertainty the expected number of events. The variation of the top-quark mass fit is assigned as a systematic uncertainty.

The effect due to the scale used to match clustered jets to partons (\ie\ jet-parton matching) is estimated with dedicated samples generated
by varying the nominal matching \pt\ thresholds  from the default of 20\GeV down to 10\GeV and up to 40\GeV.
Effects due to the definition of the renormalisation and factorisation scales used in the simulation of the signal are studied
with dedicated Monte Carlo samples
with both scales varied
by factors of 2 or $\frac{1}{2}$.

The uncertainty due to pileup is evaluated from pseudo-experiments where the total inelastic cross section used to simulate the pileup is varied 
within its uncertainty, which is estimated to be 8\%.
The uncertainties related to the parton distribution function (PDF) used to model the hard scattering of the proton-proton collisions is evaluated from pseudo-experiments for which
the distribution of $\mtop$ was obtained after varying parameters of the PDF by $\pm 1 \sigma$ with respect to their nominal values and  using the PDF4LHC prescription~\cite{Nadolsky:2008zw,cteq_2010,pdf4lhc}.
The differences found with respect to the nominal prediction are added in quadrature to obtain the total PDF uncertainty.
The uncertainties due to the underlying event~\cite{Chatrchyan:2011id} and the colour reconnection~\cite{Wicke:2008iz} are evaluated with dedicated samples.
The uncertainties due to the underlying event
are estimated by comparing two alternative \PYTHIA tunes with increased and decreased underlying event activity relative to a central tune.
The results for the top-quark mass measured in pseudo-experiments using the Perugia 2011 tune are thus compared to the Perugia 2011 mpiHi and Perugia 2011 Tevatron tunes~\cite{Skands:2010ak}.
The difference found between the two samples is taken as an estimate of the uncertainty in the modelling of the underlying event in our simulation.
The Perugia 2011 noCR tune is a variant in which colour reconnection effects are not taken into account.
The difference in the average top-quark mass, measured with and without colour reconnection effects,
is taken as the estimate for the colour reconnection systematic uncertainty.
Finally, the uncertainty due to the modelling of the signal templates by the Monte Carlo generator are studied by comparing the results of the pseudo-experiments
using the reference sample to that from a sample generated with the \textsc{powheg} generator.

\section{Measurement with the full kinematic analysis}
\label{sec:kinb}

An alternative measurement is performed using the KINb method~\cite{Chatrchyan:2011nb} and a tighter event selection.
The jet \pt is required to be at least 35\GeV and the reconstructed \MET of $\Pe \mu$ events is required to be at least 30\GeV.
These tighter requirements are expected to improve the resolution of the method.
In KINb, as in the AMWT method, the kinematic equations describing the \ttbar system
are solved many times per event for each lepton-jet combination.
The longitudinal momentum of the \ttbar system ($p_z^{\ttbar}$)
is used as the extra constraint required to solve the equations.
The jet $\pt$, the \MET direction, and the $p_z^{\ttbar}$ are varied independently according
to their resolutions in order to scan the kinematic phase space consistent with the \ttbar system.
The jet $\pt$ resolution is obtained from the data~\cite{Chatrchyan:2011ds};
the $p_z^{\ttbar}$ description, that is minimally dependent on $\mtop$, is taken from simulation.
The solution with the lowest invariant mass of the \ttbar system is accepted if the mass difference
between the top quark and antiquark masses is less than 3\GeV.
The combination of leptons and jets yielding
the largest number of solutions is chosen,
and the mass value $m_\mathrm{KINb}$
is estimated by means of a Gaussian fit to the distribution of solutions in a 50\GeV window built around the most probable value.
A key point in the method is the choice of the jets used to reconstruct the top-quark candidate,
favouring jets that have higher value of the \cPqb-tagging discriminator.
Simulations demonstrate that the proportion of events in which the jets used for the reconstruction
are correctly matched to partons from top quark decays is increased significantly
with respect to a choice based on the two jets with highest $\pt$.
Only events with solutions contribute to the
$\mtop$ measurement; in simulation,
solutions are found for 80\% of signal events
and 70\% of background events.

We use a two-component unbinned maximum-likelihood fit to the $m_\mathrm{KINb}$ distribution
to mitigate the effect of background and signal events with misreconstructed top-quark masses
and obtain an estimate of $\mtop$.
The free parameters of the likelihood are $\mtop$ and the numbers of signal and background events.
The main background contribution is from the DY events, which is estimated from data
using a template fit to the angle between the momenta of the two leptons.
Depending on $\mtop$, the signal and background templates may resemble each other;
therefore the number of background events is constrained by a Gaussian term in the likelihood function.
The parameters of signal and background templates are taken from simulation and fixed in the fit.
The signal shape is obtained with a simultaneous fit of simulated \ttbar samples to a Gaussian plus Landau function template
with parameters that are linear functions of $\mtop$.
Separate templates are used for the four samples corresponding to the same
or different flavour dileptons with one or two and more \cPqb-tagged jets.
In each category the backgrounds are added in the expected proportions.
The expected distribution from DY events is determined from data near the \Z peak
($76<M_{\ell\ell}<106~\GeV$)
for same-flavour dileptons.
From simulation, the template obtained near the \Z peak is expected to describe well DY events in the signal region.
In the case of different-flavour dileptons we estimate the contribution from DY events using a data sample of $\Z\rightarrow\mu\mu$,
by replacing the muons with fully simulated decays of $\tau$ leptons~\cite{Chatrchyan:2012vp}
and applying the event selection and top-quark mass reconstruction.
For single top quark, diboson, and other residual backgrounds the templates are taken from simulation.

\begin{figure}[htp]
\centering
\includegraphics[width=\cmsFigWidth]{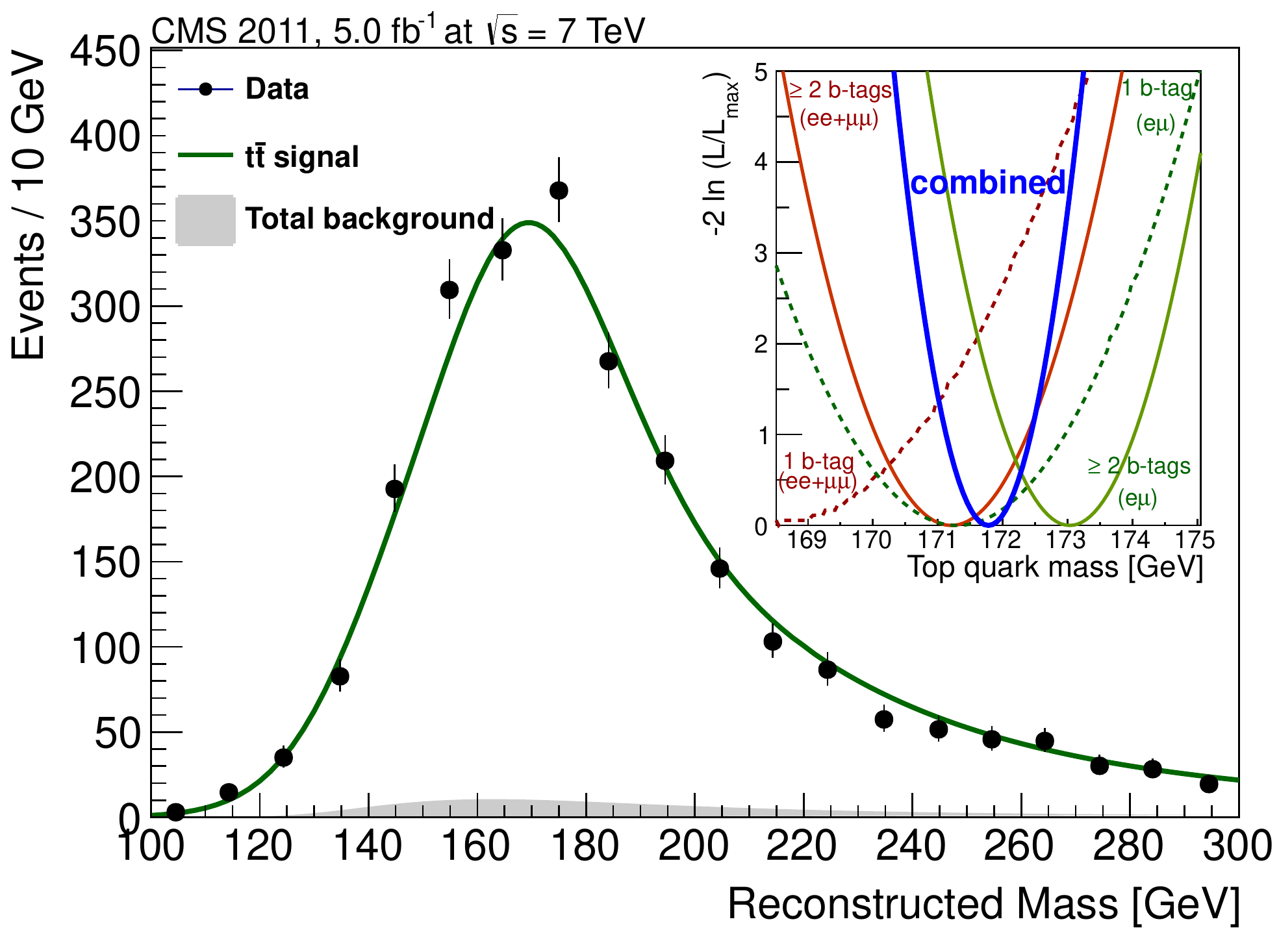}
\caption{
Result of the fit to \emu events with at least two \cPqb-tagged jets using the KINb method.
The inset shows the variation of the likelihood used to extract the top-quark mass,
in the different event categories and for all channels combined.
}
\label{fig:topmass_ll}
\end{figure}

The fit is performed separately for same- and different-lepton flavour events with either one or at least two \cPqb-tagged jets
using an unbinned likelihood method, where the inputs are the mass value returned by the KINb method in the data,
and the probability density function for signal and background.
The data in the range $100<m_\mathrm{KINb} \cmsEPJCbreak{3} <300\GeV$ is used in the fit.
Figure~\ref{fig:topmass_ll} (inset) shows the variation of $-2\ln({\cal L}/{\cal L}_{\rm max})$ as a function of $\mtop$,
for the different categories individually and for all categories combined.
For each event category the corresponding likelihood is maximised, yielding an estimate of the top-quark mass value
as well as the expected numbers of signal and background events.
The result of the fit for the category of events with the smallest
background contamination (\emu events with at least two \cPqb-tagged jets)
is shown in Figure~\ref{fig:topmass_ll}.

The expected contamination from background events and the result obtained from the fit in each category agree well.
A combined unbinned likelihood is constructed in order to extract the final measurement of $\mtop$ from data.
To minimise any residual bias resulting from the parameterisations of the signal and background $m_\mathrm{KINb}$ distributions,
pseudo-experiments are performed using simulated dilepton events generated with different $\mtop$ values.
The resulting $\mtop$ distributions are used to calibrate the parametrisation of the signal template.
We find an average bias on $\mtop$ of $0.4\pm 0.2\GeV$, which we use to correct our final value.
We assign the envelope of the residual bias (0.2\GeV) as the systematic uncertainty associated with the fit.

Other sources of systematic uncertainty are similar and fully correlated with those in the AMWT analysis.
We observe however that the KINb method is affected by larger uncertainties compared to the AMWT method,
reflecting the fact that the mass resolution
is slightly poorer.
The degradation of the resolution is related to the fact that a choice is made for the lepton-jet assignment in the event
and that there is no reweighting of the solutions found
based on any expectation for the kinematic properties,
such as polarization effects which are intrinsically modelled by Eq.~\ref{eq:probdens}.
We find no improvement in combining the AMWT and KINb given the
difference in statistical uncertainty achieved and the dominance of the correlated
systematic uncertainties. 
The KINb analysis is thus used as a cross-check, and we measure
$\mtop=171.8\pm 0.6\stat\pm 2.2\syst\GeV$,
in agreement with the AMWT measurement.

\section{Summary}
\label{sec:summary}

In summary, 
a measurement of the top-quark mass from \ttbar decays to dilepton final states is presented, 
using a data sample corresponding to an integrated luminosity of 5.0\fbinv 
recorded by the CMS experiment at $\sqrt{s} = 7\TeV$.
The measurement yields $\mtop=172.5\pm 0.4\stat \pm1.5\syst\GeV$.
An alternative measurement gives a consistent result.
With respect to the previous measurement in the dilepton channel performed by CMS on the $36\pbinv$  data collected in 2010~\cite{Chatrchyan:2011nb}, the systematic uncertainty
could be reduced substantially by improved understanding of the effect of pileup, underlying event and the uncertainty on the JES.
To date, this measurement is the most precise determination of the top-quark mass in the dilepton channel.

\section*{Acknowledgements}
{\tolerance=500
We congratulate our colleagues in the CERN accelerator departments for the excellent performance of the LHC machine. We thank the technical and administrative staff at CERN and other CMS institutes, and acknowledge support from: BMWF and FWF (Austria); FNRS and FWO (Belgium); CNPq, CAPES, FAPERJ, and FAPESP (Brazil); MES (Bulgaria); CERN; CAS, MoST, and NSFC (China); COLCIENCIAS (Colombia); MSES (Croatia); RPF (Cyprus); MoER, SF0690030s09 and ERDF (Estonia); Academy of Finland, MEC, and HIP (Finland); CEA and CNRS/IN2P3 (France); BMBF, DFG, and HGF (Germany); GSRT (Greece); OTKA and NKTH (Hungary); DAE and DST (India); IPM (Iran); SFI (Ireland); INFN (Italy); NRF and WCU (Korea); LAS (Lithuania); CINVESTAV, CONACYT, SEP, and UASLP-FAI (Mexico); MSI (New Zealand); PAEC (Pakistan); MSHE and NSC (Poland); FCT (Portugal); JINR (Armenia, Belarus, Georgia, Ukraine, Uzbekistan); MON, RosAtom, RAS and RFBR (Russia); MSTD (Serbia); SEIDI and CPAN (Spain); Swiss Funding Agencies (Switzerland); NSC (Taipei); TUBITAK and TAEK (Turkey); STFC (United Kingdom); DOE and NSF (USA).

Individuals have received support from the Marie-Curie programme and the European Research Council (European Union); the Leventis Foundation; the A. P. Sloan Foundation; the Alexander von Humboldt Foundation; the Austrian Science Fund (FWF); the Belgian Federal Science Policy Office; the Fonds pour la Formation \`a la Recherche dans l'Industrie et dans l'Agriculture (FRIA-Belgium); the Agentschap voor Innovatie door Wetenschap en Technologie (IWT-Belgium); the Council of Science and Industrial Research, India; the Compagnia di San Paolo (Torino); and the HOMING PLUS programme of Foundation for Polish Science, cofinanced from European Union, Regional Development Fund.\par}

\bibliography{auto_generated}

\cleardoublepage \appendix\section{The CMS Collaboration \label{app:collab}}\begin{sloppypar}\hyphenpenalty=5000\widowpenalty=500\clubpenalty=5000\textbf{Yerevan Physics Institute,  Yerevan,  Armenia}\\*[0pt]
S.~Chatrchyan, V.~Khachatryan, A.M.~Sirunyan, A.~Tumasyan
\vskip\cmsinstskip
\textbf{Institut f\"{u}r Hochenergiephysik der OeAW,  Wien,  Austria}\\*[0pt]
W.~Adam, E.~Aguilo, T.~Bergauer, M.~Dragicevic, J.~Er\"{o}, C.~Fabjan\cmsAuthorMark{1}, M.~Friedl, R.~Fr\"{u}hwirth\cmsAuthorMark{1}, V.M.~Ghete, J.~Hammer, N.~H\"{o}rmann, J.~Hrubec, M.~Jeitler\cmsAuthorMark{1}, W.~Kiesenhofer, V.~Kn\"{u}nz, M.~Krammer\cmsAuthorMark{1}, I.~Kr\"{a}tschmer, D.~Liko, I.~Mikulec, M.~Pernicka$^{\textrm{\dag}}$, B.~Rahbaran, C.~Rohringer, H.~Rohringer, R.~Sch\"{o}fbeck, J.~Strauss, A.~Taurok, W.~Waltenberger, G.~Walzel, E.~Widl, C.-E.~Wulz\cmsAuthorMark{1}
% ...
\vskip\cmsinstskip
\dag:~Deceased\\
1:~~Also at Vienna University of Technology, Vienna, Austria\\
% ...

\end{sloppypar}
\end{document}